\documentclass[]{pasj01}
\usepackage{graphicx}
\usepackage{natbib}
\usepackage{aas_macros}
\usepackage{ulem}
\def\Vec#1{\mbox{\boldmath $#1$}}
\def\D#1#2{\frac{d #1}{d #2}}
\def\DD#1#2{\frac{d^2 #1}{d #2^2}}

\def\P#1#2{\frac{\partial #1}{\partial #2}}
\def\PP#1#2{\frac{\partial^2 #1}{\partial #2^2}}

\def\k0{\kappa_0}

\begin{document}
\Received{2015 February 27}{}
\Accepted{2015 March 27}{}

 \title{
Appearance of the prolate and
the toroidal magnetic field dominated stars \\ -- Analytic approach}

\author{Kotaro \textsc{Fujisawa}\altaffilmark{1}}
\altaffiltext{1}{Advanced Research Institute for Science and Engineering, 
    Waseda University, 3-4-1 Okubo, Shinjuku-ku, Tokyo    169-8555, Japan}
\email{fujisawa@heap.phys.waseda.ac.jp}

\author{Yoshiharu \textsc{Eriguchi}\altaffilmark{2}}
\altaffiltext{2}{Department of Earth Science and Astronomy,
	 Graduate School of Arts and Sciences, University of Tokyo, 
 	 Komaba, Meguro-ku, Tokyo 153-8902, Japan} 


\maketitle

\begin{abstract}
We have analyzed magnetized equilibrium states and showed 
a condition for appearance of  the prolate and  
the toroidal magnetic field 
dominated stars by analytic approaches.
Both observations and numerical stability analysis support that 
the magnetized star would have 
the prolate and the 
large internal toroidal 
magnetic fields. 
In this context, many investigations concerning magnetized equilibrium states 
have tried to obtain 
 the prolate and 
the toroidal dominant solutions, 
but many of them have failed to obtain such 
configurations. Since the Lorentz force 
is a cross product of current density and 
magnetic field, 
 the prolate shaped configurations and the 
large toroidal magnetic fields 
in stars require a special relation between current density 
and the Lorentz force. 
We have analyzed simple analytical solutions and found that 
the prolate and the toroidal dominant configuration require 
non force-free toroidal current density
that flows in the opposite direction with respect to 
the bulk current within the star.
Such current density results in the Lorentz force 
which makes the stellar 
shape prolate. Satisfying this special relation 
between the current density
and the Lorentz force is a key for appearance 
of the prolate and the toroidal magnetic field 
dominated magnetized star.
\end{abstract}


\section{Introduction}

Anomalous X-ray Pulsars and 
Soft-Gamma-ray-Repeaters (SGR's)
are considered as special classes of neutron stars, 
i.e. magnetars (\citealt{Thompson_Duncan_1995}). 
According to observations of rotational periods and 
their time derivatives, magnitudes of global dipole magnetic 
fields of magnetars reach about $10^{14-15}$G.  Recently, 
however, SGR's with weak dipole magnetic fields have
been found (\citealt{Rea_et_al_2010,Rea_et_al_2012}).
Their observational characteristics are very similar 
to those of ordinary SGR's but their global dipole 
magnetic fields are much weaker than those of 
ordinary magnetars. It might be explained by a 
possibility that such SGR's with small magnetic fields
hide large toroidal magnetic fields under their
surfaces and drive their activities by their internal 
toroidal magnetic energy (\citealt{Rea_et_al_2010}). 
Recent X-ray observation of magnetar 4U 0142+61
also implies the presence of large toroidal magnetic fields
 and the possibility of prolate-shaped neutron star
 (\citealt{Makishima_Enoto_et_al_2014}).
By considering possible growth of magnetic fields of 
magnetars during proto-magnetar phases,
strong differential rotation within proto-magnetars
would amplify their toroidal magnetic fields 
(\citealt{Duncan_Thompson_1992}; \citealt{Spruit_IAU259}).
Therefore, it would be natural that some magnetars 
sustain large toroidal magnetic fields inside.

The large toroidal fields are required from 
the stability analyses of magnetic fields. 
Stability analyses have shown that stars with
purely poloidal fields or purely toroidal fields 
are unstable (\citealt{Markey_Tayler_1973}; \citealt{Tayler_1973}). 
Stable magnetized stars should have both poloidal and 
toroidal magnetic fields. Moreover, the toroidal magnetic
field strengths of the stable magnetized stars have been
considered to be comparable with those of poloidal 
components (\citealt{Tayler_1980}).
However, we have not yet known the exact stability 
condition and stable magnetic field configurations, 
because it is too difficult to carry out stability 
analyses of stars with both poloidal and toroidal magnetic 
fields.

Nevertheless, stabilities of magnetic fields have been
studied by performing dynamical simulations.
\cite{Braithwaite_Spruit_2004} showed that twisted-torus magnetic field 
structures are stable magnetic field configurations on dynamical timescale.
Stabilities of purely toroidal magnetic field configurations or purely poloidal
magnetic field configurations have been studied in the Newtonian framework 
(\citealt{Braithwaite_2006,Braithwaite_2007}) and
in the full general relativistic framework (\citealt{Kiuchi_Yoshida_Shibata_2011}; 
\citealt{Lasky_et_al_2011}; \citealt{Ciolfi_et_al_2011};
\citealt{Ciolfi_Rezzolla_2012}).
\cite{Braithwaite_2009} and \cite{Duez_Braithwaite_Mathis_2010}
have found a stability criterion of the twisted-torus magnetic 
fields. It could be expressed as
\begin{eqnarray}
 \alpha \frac{{\cal M}}{|W|} < \frac{{\cal M}_p}{{\cal M}} \leq 0.8,
\end{eqnarray}
where ${\cal M} / |W|$ is the ratio of the total 
magnetic energy to the gravitational energy. 
${\cal M}_p / {\cal M}$ is the 
ratio of the poloidal magnetic field energy 
to the total magnetic field energy.
$\alpha$ is a certain dimensionless factor 
of order of 10 for main-sequence stars
and of order $10^3$ for neutron stars. 
The ratio of ${\cal M} / |W|$ is a small
value ($\sim 10^{-5}$) even for magnetars.
Therefore, the criterion becomes
\begin{eqnarray}
 0.2 \leq  \frac{{\cal M}_t}{\cal M} \lesssim 0.99,
\end{eqnarray}
where ${\cal M}_t$ is the toroidal magnetic field energy.
Stellar magnetic fields would be stable even for
toroidal magnetic field dominated configurations. 
Therefore, it is very natural that the 
toroidal magnetic field strength of the stable stationary 
magnetized stars are comparable with or larger than those of poloidal component. 

Until recently, however, almost all numerically
obtained equilibrium configurations for 
stationary and axisymmetric stars have only small fractions
of toroidal magnetic fields, typically 
${\cal M}_t / {\cal M} \sim 0.01$, even for 
twisted-torus magnetic field configurations in
the Newtonian gravity (\citealt{Tomimura_Eriguchi_2005};
\citealt{Yoshida_Eriguchi_2006}; \citealt{Yoshida_Yoshida_Eriguchi_2006};
\citealt{Lander_Jones_2009}; \citealt{Lander_Andersson_Glampedakis_2012};
\citealt{Fujisawa_Yoshida_Eriguchi_2012}; \citealt{Lander_2013a, Lander_2014};
\citealt{Bera_Bhattacharya_2014}; \citealt{Armaza_et_al_2014}),
 in general relativistic perturbative solutions 
(\citealt{Ciolfi_et_al_2009}; \citealt{Ciolfi_et_al_2010}),
 and general relativistic non-perturbative solutions 
under both simplified relativistic gravity (\citealt{Pili_et_al_2014})
and fully relativistic gravity (\citealt{Uryu_et_al_2014}).
All of them  do not satisfy the stability criterion mentioned 
above.

On the other hand, there appeared several works 
which have successfully obtained the stationary
states with strong toroidal magnetic fields
by applying special boundary conditions.
\cite{Glampedakis_Andersson_Lander_2012} obtained
strong toroidal magnetic field 
models imposing surface currents
on the stellar surface  as their boundary 
condition. \cite{Duez_Mathis_2010} imposed the 
boundary condition that the magnetic flux on 
the stellar surface should vanish.
Since the magnetic fluxes of their models are
zero on the stellar surfaces, all the magnetic 
field lines are confined within the stellar
surfaces. They obtained configurations
with strong toroidal magnetic fields
which are essentially the same as those of 
classical works by \cite{Prendergast_1956}, 
\cite{Woltjer_1959a, Woltjer_1959b, Woltjer_1960} and \cite{Wentzel_1960, Wentzel_1961}
and recent general relativistic works by \cite{Ioka_Sasaki_2004} and
\cite{Yoshida_Kiuchi_Shibata_2012}.

It is very recent that \cite{Fujisawa_Eriguchi_2013} 
have found and shown that the strong toroidal magnetic fields within the stars
require 
the non force-free current or surface current which flows in the 
opposite direction with respect to the bulk current within the star.
Such oppositely flowing currents can sustain large 
toroidal magnetic fields in magnetized stars. 
It is also very recent that \cite{Ciolfi_Rezzolla_2013} have 
obtained stationary states of twisted-torus 
magnetic field structures with  very  strong  
toroidal  magnetic  fields using a special choice 
for the toroidal current. Their toroidal currents 
contain oppositely flowing current 
components and result in the large toroidal magnetic fields,
although their paper does not explain the physical meanings  
for appearances of such oppositely
 flowing toroidal currents.
They also did not show clear conditions for the
appearance of the toroidal magnetic field dominated stars.

On the other hand, strong poloidal magnetic fields make 
stellar shape oblate one (e.g. \citealt{Tomimura_Eriguchi_2005}), 
but the strong toroidal magnetic field
tends to stellar shape prolate one (\citealt{Haskell_et_al_2008}; 
\citealt{Kiuchi_Yoshida_2008}; \citealt{Lander_Jones_2009};
\citealt{Ciolfi_Rezzolla_2013})
Since the Lorentz force is a cross product of the 
current density and magnetic field, 
such Lorentz force requires a special relation 
between the magnetic fields and the current density.
At the same time, the large toroidal magnetic fields in stars also 
need a special relation between current density and Lorentz force.
The oppositely flowing toroidal current density is a key 
to reveal these relations and a condition for appearance of the 
toroidal magnetic field dominated stars.

We analyze magnetic field configurations 
and consider the special relations in this paper. 
We find a condition for the 
appearance of the toroidal magnetic field dominated stars,
which was not described in our previous work (\citealt{Fujisawa_Eriguchi_2013}).
In order to show the relations and condition clearly, simplified  analytical models are solved  
and we show examples of the prolate configurations 
and the large toroidal magnetic fields within stars.
This paper is organized as follows. 
The formulation and basic equations are shown 
in Sec. 2. We present analytic solutions and the relations.
We also explain the important role of 
oppositely flowing components of the $\kappa$ 
currents
for appearance of the prolate shapes and the 
presence of the large toroidal 
magnetic fields using the relations in Sec. 3. 
Discussion and conclusions follow in Sec. 4.
In Appendix \ref{App:N1} the deformation of stellar shape and
the gravitational potential perturbation are briefly
summarized.

\section{Formulation and basic equations}
\label{Sec:Formulations_Methods} 

Stationary and axisymmetric magnetized barotropic
stars without rotation and meridional flows are analyzed 
in this paper.

Some authors have claimed that there do not
exist dynamically stable barotropic 
magnetized stars (e.g. \citealt{Mitchell_et_al_2015}
), but in our opinion, their arguments
should be applied only to {\it isentropic
barotropes}. The magnetized barotropic stars
are well defined concepts apart from
the thermal stability or convective stability
due to the entropy distributions and/or due to
the chemical composition distributions. Thus we will
investigate {\it mechanical equilibrium states}
of traditionally defined barotropes 
(e.g., \citealt{Chandrasekhar_Prendergast_1956};
\citealt{Prendergast_1956}) in this paper.

Since for such configurations the basic 
equations and basic relations are
shown, e.g., in \cite{Tomimura_Eriguchi_2005} 
and \cite{Fujisawa_Eriguchi_2013},
we  show the basic equations
and basic relations briefly. 

The stationary condition for the configurations
mentioned above can be expressed as:
\begin{eqnarray}
  \label{Eq:stationary-condition}
 \int \frac{dp}{\rho} = - \phi_g + \int \mu(\Psi)\, d \Psi  + C,
\end{eqnarray}
where $\rho$, $p$, $\phi_g$ and $C$ are the density, 
the pressure, the gravitational potential of 
the star and an integral constant, respectively.
$\Psi$ is the magnetic flux function.
$\mu$ is an arbitrary function of $\Psi$. The magnetic flux is governed by
\begin{eqnarray}
 \Delta^* \Psi= - 4\pi r \sin \theta \frac{j_\varphi}{c}.
\label{Eq:GS}
\end{eqnarray}
where
\begin{eqnarray}
 \Delta^* = \left(\PP{}{r} + \frac{1}{r^2} \PP{}{\theta} 
- \frac{1}{r^2}\frac{\cos \theta}{\sin \theta} \P{}{\theta} \right) \ ,
\end{eqnarray}
and $j_{\varphi}$ is a $\varphi$-component,
i.e. the toroidal component, of the 
current density. The spherical coordinates
$(r, \theta, \varphi)$ are used.

From the integrability condition of the equation of 
motion, the axisymmetry and the stationarity, 
the following relations are derived:
\begin{eqnarray}
 \frac{\Vec{j}}{c} = \frac{1}{4\pi}\D{\kappa}{\Psi} \Vec{B} + \rho r \sin \theta 
\mu(\Psi) \Vec{e}_\varphi,
\end{eqnarray}
\begin{eqnarray}
 \kappa = \kappa(\Psi) \ ,
\end{eqnarray}
where $\Vec{j}$ and $\Vec{B}$ are the current density
and the magnetic field, respectively, and $\kappa$ is
another arbitrary function of $\Psi$.
It would be helpful to note that the above relation 
for $\kappa$ was found by \cite{Mestel_1961} and 
\cite{Roxburgh_1966}.
Although $\kappa(\Psi)$ is exactly a function
of the magnetic flux function only in stationary and 
axisymmetric system (\citealt{Braithwaite_2009}), 
\cite{Braithwaite_2008} 
showed that the function $\kappa(\Psi)$ during 
the dynamical evolution of magnetized 
configurations is nearly conserved 
even for non-axisymmetric systems.

Since the function $\kappa$ and the $\varphi$-component
of the magnetic field $B_{\varphi}$ is related as
\begin{eqnarray}
 \kappa(\Psi)  = r \sin \theta B_{\varphi} \ ,
\end{eqnarray}
the toroidal current density can be expressed 
as:
\begin{eqnarray}
 \frac{j_\varphi}{c} = \frac{1}{4\pi} 
\frac{\kappa(\Psi) \kappa'(\Psi)}{r \sin \theta} + \rho r\sin \theta 
\mu(\Psi).
\end{eqnarray}

Under our assumption that the magnetic field energy is small
compared to the gravitational energy (${\cal M}/ |W| < 10^{-5}$) 
in this paper,  the influence of the magnetic fields can 
be treated as a small perturbation to a 
spherical star. Therefore, we assume that the 
stellar configurations are sphere and that the 
density profile depends only on $r$, i.e. 
$\rho = \rho(r)$. For such situations, we can 
obtain analytical solutions easily.
Noted that self-consistent approaches such as \cite{Tomimura_Eriguchi_2005}, 
would reveal some differences in the magnetic field solutions.
In self-consistent approaches, we need to calculate both 
magnetic fields and matter equations iteratively. 
The stellar shape is no longer sphere and the stellar 
configuration affects the magnetic field configuration. 
However, our result in this paper is simple and 
might be important for both perturbative and self-consistent approaches.

\section{Spherical models with weak magnetic fields}
\label{Sec:Results} 

Our aim in this paper is investigating the condition for 
appearance of the toroidal magnetic field dominated star analytically.
Noted that our solutions of $\Psi$ themselves are classical and not new ones, 
but we use the solutions in order to show the special condition clearly. 

\subsection{Green's function approach and analytic solutions}

We follow mostly the formulation of the classical works  
(\citealt{Chandrasekhar_Prendergast_1956}; 
\citealt{Prendergast_1956}; \citealt{Woltjer_1959a}; \citealt{Woltjer_1959b, Woltjer_1960};
\citealt{Wentzel_1960, Wentzel_1961}) 
and the recent analytical works (\citealt{Broderick_Narayan_2008};
\citealt{Duez_Mathis_2010}; \citealt{Fujisawa_Eriguchi_2013}).
In order to obtain analytical solutions, 
we choose the functional forms as follows:
\begin{eqnarray}
 \mu(\Psi) = \mu_0,
\label{Eq:mu0}
\end{eqnarray}
\begin{eqnarray}
 \kappa(\Psi) = \kappa_0 \Psi \ ,
\label{Eq:I0}
\end{eqnarray}
where $\mu_0$ and $\kappa_0$ are two constants.
It should be noted that these functional forms always lead to non-zero 
surface currents unless magnetic fields are confined inside the star.
The surface current induces a Lorentz force at the stellar surface (\citealt{Lander_Jones_2012}).
It would be unphysical because the Lorentz force need to be balanced by other physics 
such as a crust of the neutron star (e.g. \citealt{Fujisawa_Kisaka_2014}). 
The models relying on a surface current might not be physically realistic. 
We emphasize that we are not asserting that surface currents themselves 
are necessarily significant in real stars.
The surface current simply provides a mathematically convenient way
of describing analytical solutions easily.

By using these functional forms, the toroidal current density can be expressed as:
\begin{eqnarray}
 \frac{j_\varphi}{c} = \frac{1}{4\pi} \frac{\kappa_0^2 \Psi}{r\sin \theta} + \mu_0 \rho (r) r \sin\theta . 
\label{Eq:j_phi}
\end{eqnarray}
We name the first term $\kappa$ current $j^{\kappa_{\varphi}}$
 (force-free) term and the second term 
$\mu$ current $j^{\mu}_{\varphi}$ (non force-free) term, respectively (\citealt{Fujisawa_Eriguchi_2013}).
Then, the equation for the magnetic flux becomes 
as follows:
\begin{eqnarray}
\Delta^* \Psi + \kappa_0^2 \Psi  = - 4 \pi \mu_0 \rho(r) r^2 \sin^2 \theta. 
\end{eqnarray}
It should be noted that this is 
a linear equation for the magnetic flux function 
with the inhomogeneous term which contains
multipoles less than the quadrupole.
If we impose the boundary condition 
$\Psi = 0$ at the center of the star, 
$\Psi$ is described as follows (\citealt{Duez_Mathis_2010}; \citealt{Fujisawa_Eriguchi_2013}):
\begin{eqnarray}
 \frac{\Psi}{\sin^2 \theta} &=& K \kappa_0 r j_1 \left(\kappa_0 r \right) \nonumber \\ 
 &-& 4\pi \mu_0 \kappa_0  \Bigg\{ 
r j_{1} \left(\kappa_0 r \right) \int_{r}^{r_s=1} y_{1}
\left(\kappa_0 r' \right) \rho(r') r'^3 \, dr' \nonumber \\
 &+& r y_{1} \left(\kappa_0 r \right) \int_0^{r} j_{1}
\left(\kappa_0 r' \right) \rho(r') r'^3 \, dr'
\Bigg\}.
\label{Eq:Psi_anal}
\end{eqnarray}
where we set the stellar radius $r_s = 1$ in this paper.
$j_1$ and $y_1$ are the spherical Bessel functions 
of the first kind and the second kind, respectively
and $K$ is a coefficient which is determined 
by a boundary condition of $\Psi$ at the surface.
According to the $\theta$-dependency of the 
inhomogeneous term,
we search for solutions of the following form:
\begin{eqnarray}
a (r)\sin^2 \theta \equiv \Psi (r,\theta).
\end{eqnarray}
Therefore we obtain the solution for $a(r)$ by imposing 
the boundary condition at the surface and integrating equation (\ref{Eq:Psi_anal}).

In this paper, we treat spherical polytropes with
the polytropic indices $N = 0$ and $N = 1$.
As for the configurations of the magnetic fields,
we choose two types: (1) closed field models  
 (e.g. \citealt{Duez_Mathis_2010}) and (2) open field
models (e.g. \citealt{Broderick_Narayan_2008}).
For closed field models, since all magnetic field 
lines are closed and confined within the
star, the magnetic flux must vanish at the 
stellar surface as follows:
\begin{eqnarray}
 a(r_s) = 0 \ .
\label{Eq:closed}
\end{eqnarray}
For open field models, since the poloidal magnetic 
field lines must continue smoothly through
 the stellar surfaces into the outside. 
the boundary condition can be expressed as:
\begin{eqnarray}
 a(r_s) = -\D{a(r)}{r} \Bigg|_{r=r_s} \ .
\label{Eq:open}
\end{eqnarray}
The density profiles are 
\begin{eqnarray}
 \rho(r) = \rho_c,
\end{eqnarray}
for $N=0$ polytrope and
\begin{eqnarray}
 \rho(r) = \rho_c \frac{\sin (\pi r)}{\pi r}.
\end{eqnarray}
for $N=1$ polytrope and  $\rho_c$ is the central density. 
We can obtain four different analytical solutions 
according to four different situations.
We name them as $a_{0C}(r)$ ($N=0$ with closed fields), 
$a_{0O}(r)$ ($N=0$ with open fields), 
$a_{1C}(r)$ ($N=1$ with closed fields) and 
$a_{1O}(r)$ ($N=1$ with open fields).

Since the poloidal magnetic field lines are continuous
smoothly at the surfaces for open field models, 
their external solutions $a^{ex}(r)$ must be expressed as:
\begin{eqnarray}
 a^{ex}(r) = \frac{a(r_s)}{r} \ .
\label{Eq:external}
\end{eqnarray}

\begin{figure*}
 \begin{center}
  \includegraphics[width=8cm]{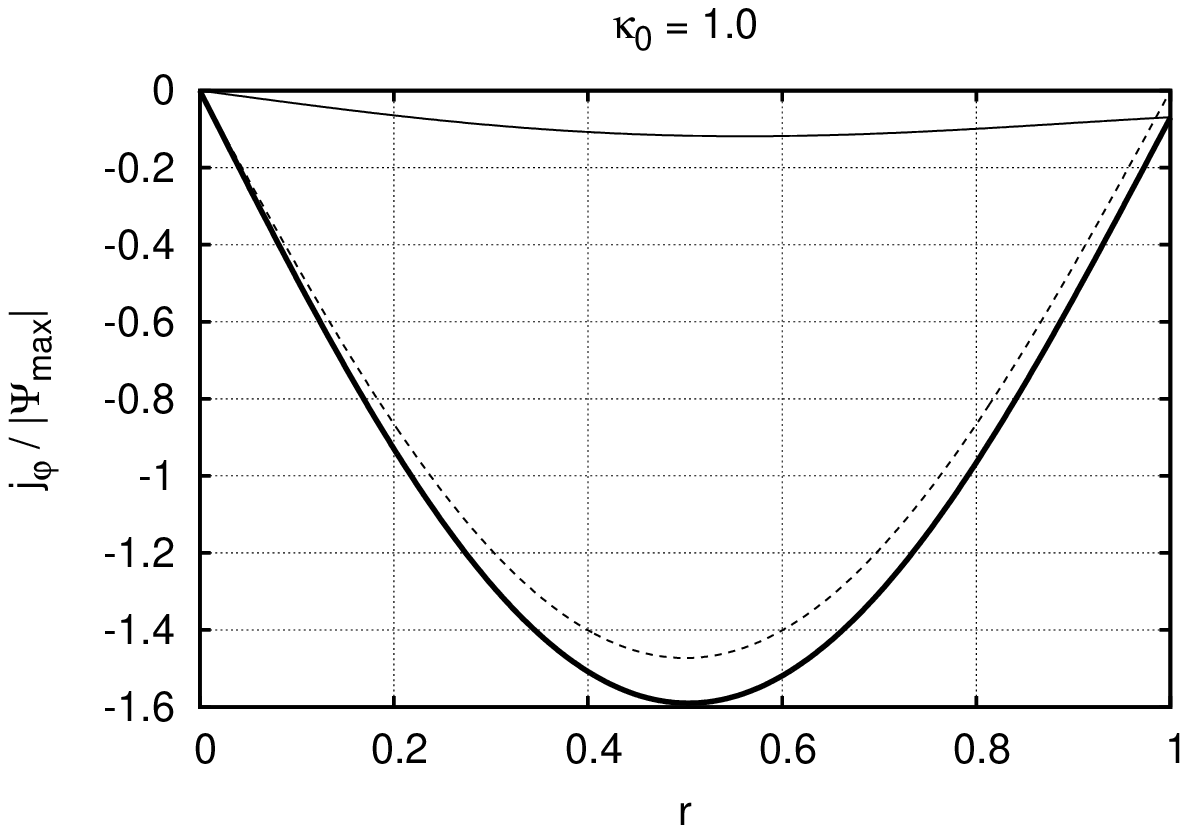}
 \includegraphics[width=8cm]{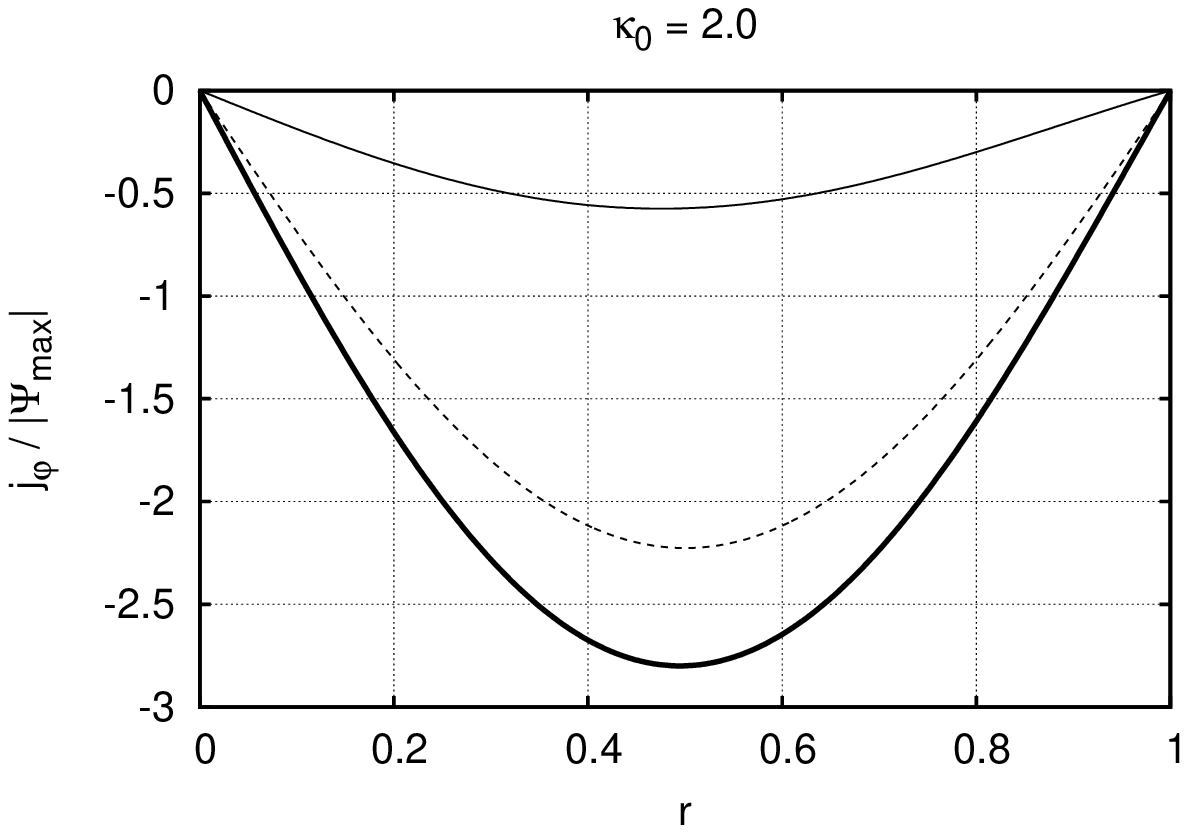}

  \includegraphics[width=8cm]{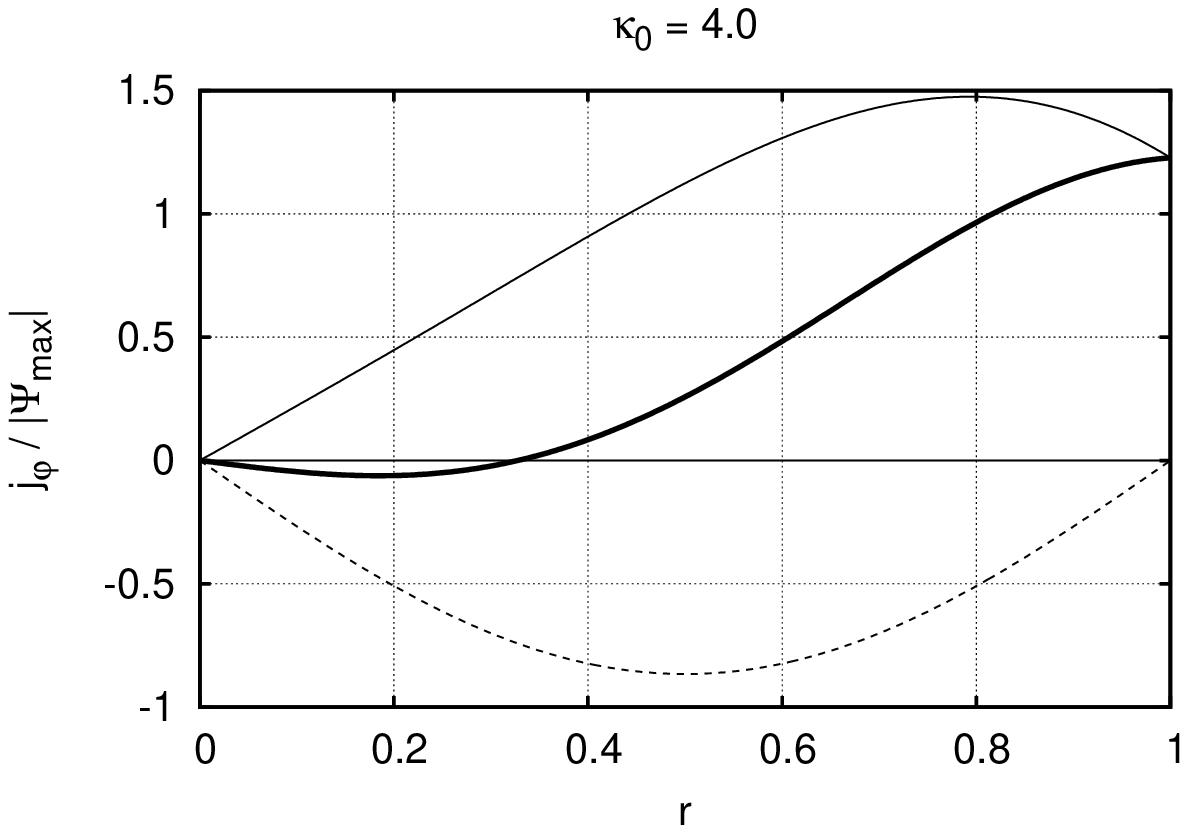}
  \includegraphics[width=8cm]{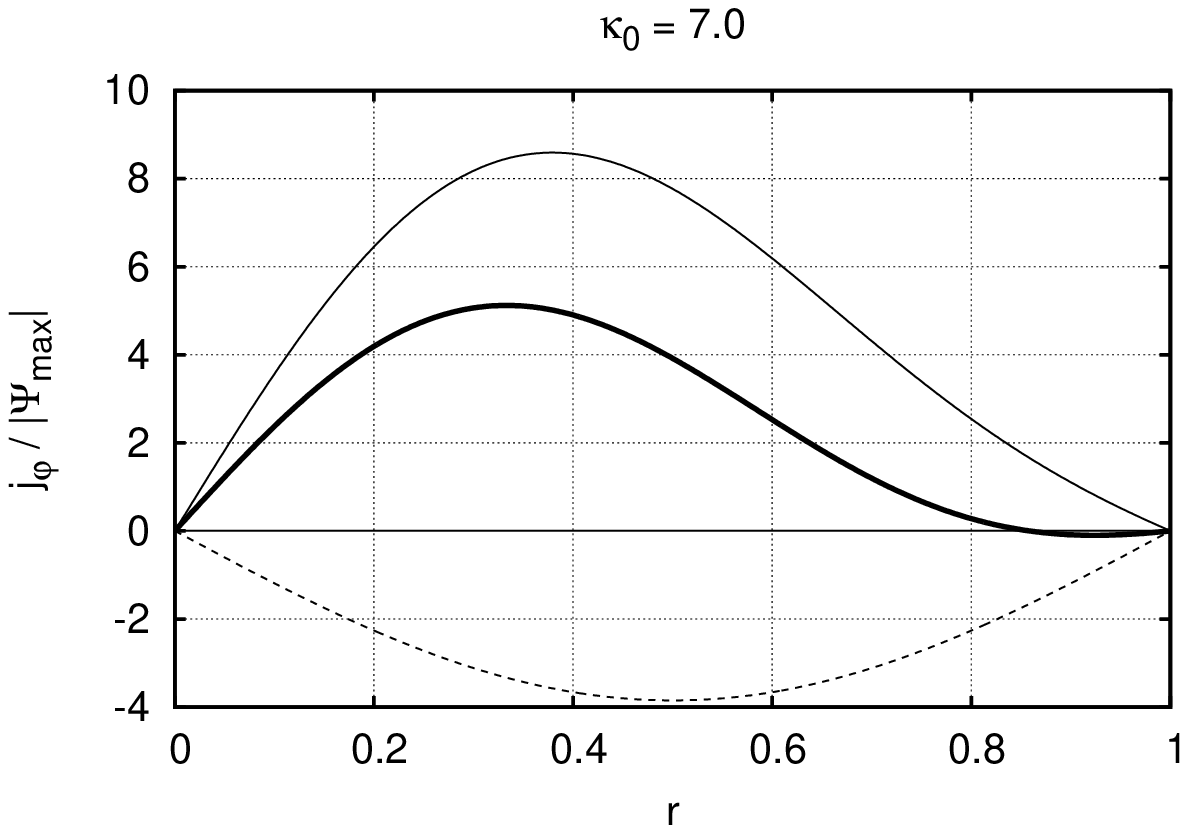}

 \end{center}

\caption{Distributions of the toroidal current density 
normalized by the maximum strength of 
$|\Psi_{\max}|$ are shown along the equatorial plane. 
Curves with different types denote the behaviors 
of the total toroidal current density, $j_\varphi / c$,
(thick solid line), the toroidal $\kappa_0$ current
density  (thin solid line) and the toroidal $\mu_0$ 
current density (thin dotted line).
We set $\mu_0 = -1$ in order to plot these distributions. 
Left panels show the profiles of solution $a_{1O}$ with $\kappa_0 = 1.0$ and 4.0
and right panels show those of solution $a_{1C}$ with $\kappa_0 = 2.0$ and 7.0.}

\label{Fig:j_r}
\end{figure*}

 It should be  noted that
 the poloidal magnetic fields for closed field models are 
 discontinuous at the
 surfaces except for solutions with special 
 values of $\kappa_0$, i.e. eigen solutions with
 corresponding eigenvalues
 (\citealt{Broderick_Narayan_2008}; \citealt{Duez_Mathis_2010}; \citealt{Fujisawa_Eriguchi_2013}).
 Therefore, these non-eigen configurations have
 toroidal  surface currents. On the other hand, the toroidal magnetic fields for open field models are always 
 discontinuous because of the choice of the functional form of $\kappa$ (equation \ref{Eq:I0}).
 It implies that the open field configurations have non-zero 
 poloidal surface currents.

 For non-eigen solutions, the 
toroidal surface current density can be expressed as follows:
 \begin{eqnarray}
   \frac{j_{\varphi,sur} (\theta)}{c} &=& 
   \frac{1}{4\pi} (B_\theta^{ex} - B_\theta^{in}  ) \Bigg|_{r=r_s} \nonumber \\ 
  &=& \frac{1}{4\pi r_s \sin \theta} \left( \P{\Psi^{ex}}{r} - \P{\Psi^{in}}{r}\right) \Bigg|_{r=r_s} \nonumber \\
 &=& \frac{\sin \theta}{4\pi r_s } \left( \D{a^{ex}}{r} - \D{a^{in}}{r}\right) \Bigg|_{r=r_s} \nonumber \\
 &=& j_0 \sin \theta,
  \label{Eq:j_0}
 \end{eqnarray}
 where superscript $^{in}$ denotes an internal solution 
and $j_0$ is a coefficient of the surface current density.

Analytic solutions are obtained by fixing a boundary condition and 
integrating equation (\ref{Eq:Psi_anal}).
Four different inner solutions $(0 \leq r \leq 1)$ can be
obtained according to four different
situations as follows:
\begin{eqnarray}
 a_{0C}(r) = 4\pi \mu_0 \rho_c \left[\frac{\sin(\kappa_0 r) - \kappa_0 r \cos (\kappa_0 r)}
{r \kappa_0^2  (\sin \kappa_0 - \kappa_0 \cos \kappa_0) }  - \frac{r^2}{\kappa_0^2} \right],
\end{eqnarray}
\begin{eqnarray}
 a_{0O} (r) = 4\pi \mu_0 \rho_c \left[
 \frac{3\{\sin(\kappa_0 r) - \kappa_0 r \cos(\kappa_0 r)   \}}{r \kappa_0^4 \sin \kappa_0 }
 - \frac{r^2}{\kappa_0^2}
\right],
\end{eqnarray}
\begin{eqnarray}
 a_{1C}(r) =& \frac{\mu_0 \rho_c}{r(\kappa_0^2 - \pi^2)^2} 
\Bigg[
\frac{8\pi \{ \sin (\kappa_0 r) - \kappa_0 r \cos(\kappa_0 r)\}}{\sin \kappa_0 - \kappa_0 \cos \kappa_0} \nonumber \\
-& \left\{ \left( 4 \kappa_0^2 -  4\pi^2 \right)r^2 + 8 \right\} \sin(\pi r)
+ 8 \pi r \cos (\pi r)
\Bigg],
\end{eqnarray}
\begin{eqnarray}
a_{1O}(r) &=  \frac{\mu_0 \rho_c}{r (\kappa_0^2 - \pi^2)^2} \Bigg[
\frac{(4\pi^3 - 4\pi \kappa_0^2) \{\sin (\kappa_0 r) - \kappa_0 r \cos(\kappa_0 r)\}}{\kappa_0^2 \sin \kappa_0} \nonumber \\
&- \left\{  (4 \kappa_0^2 - 4\pi^2 )r^2 + 8 \right\} \sin (\pi r) + 8 \pi r \cos (\pi r)
\Bigg] \ .
\end{eqnarray}
The open field models ($a_{0O}$ and $a_{1O}$) continue 
to the external solutions ($r \geq 1$)
expressed by equation (\ref{Eq:external}).
Here it would be helpful to explain several different 
kinds of characteristic solutions.

First, for $a_{1C}(r)$ solutions there appears
a singular solution at $\kappa_0 = \pi$
(\citealt{Haskell_et_al_2008}), while
the solution $a_{0C}$ is not singular at 
$\kappa_0 = \pi$ (\citealt{Fujisawa_Eriguchi_2013}).

Second, although most solutions are accompanied by
surface currents, some special solutions have
no surface currents. We call such solutions without
surface currents as eigen solutions and the values of
$\kappa_0$ as eigenvalues. 

Third, there appear many eigen solutions as the value of $\kappa_0$ exceeds the first 
eigenvalue. We call those eigen solutions as higher-order eigen solutions
(see figures in \citealt{Broderick_Narayan_2008}; 
\citealt{Duez_Mathis_2010}; \citealt{Yoshida_Kiuchi_Shibata_2012}). 
Those solutions appear when the value of 
$\kappa_0$ exceeds the first eigenvalue of $\kappa_0$
for each situation.

Fourth, special solutions with different polytropic indices
come to coincide with each other. In other words,
those solutions do not depend on the matter distributions.
As seen from the expression for the current density, the 
contribution from the $\mu$ current term needs
to disappear. It implies that those solutions are
determined only by the $\kappa$ current. Since the 
$\kappa$ currents do not contribute to the
Lorentz force, these solutions are called as
the force-free solutions (\citealt{Wentzel_1961}). The force-free solution
is expressed by the following form:
\begin{eqnarray}
 a_{ff} (r) =  \frac{K}{\kappa_0 r} \left\{ \sin (\kappa_0 r) - \kappa_0 r \cos(\kappa_0 r)\right\}.
\end{eqnarray}
The solution becomes force-free when
$\kappa_0 \sim 4.49$ and 7.73 for closed models
and when $\kappa_0 = \pi$ and $2 \pi$ for open field models.
(\citealt{Broderick_Narayan_2008}; \citealt{Fujisawa_Eriguchi_2013}).

The toroidal  surface current in equation (\ref{Eq:j_0}) vanishes when the $\kappa_0$ is
eigenvalue, i.e., for eigen solutions.
The lowest eigenvalues are  $\kappa_0 \sim 5.76$ for $a_{0O}$, $\sim 7.42$ for $a_{1c}$, 
$\sim 5.76$ for $a_{0O}$ and $\kappa_0 \sim 4.66$ for $a_{1O}$.
Hereafter, we focus on solutions with $\kappa_0$ less  than
the lowest eigenvalue. However, 
our analyses and results could be general 
and would be valid even when the configurations are higher-order eigen solutions.

In figure \ref{Fig:j_r}, distributions of the normalized 
$j_\varphi /c$ (thick solid line), the $\kappa$ current 
(thin solid line) and the $\mu$ current term 
(dashed line) along the equatorial plane are
shown for solutions of $a_{1O}$ (with $\kappa_0 = 1.0, 4.0$, 
smaller and larger
 than force-free $\kappa_0$, respectively) and 
$a_{1C}$ (with $\kappa_0 = 2.0, 7.0$,
  smaller and  larger than force-free $\kappa_0$, respectively.)
We have fixed $\mu_0 = -1$ following 
\cite{Fujisawa_Eriguchi_2013} 
in order to plot these curves. 

As seen in upper  panels in figure \ref{Fig:j_r}, 
directions (signs) of the $\mu$ current, i.e. non force-free current,
and the $\kappa$ current, i.e. force-free current, 
are the same for solutions with smaller $\kappa_0$.
By contrast, for solutions with larger $\kappa_0$
(lower panels in figure \ref{Fig:j_r}), 
the $\mu$ current  flows oppositely to the $\kappa$ current 
(\citealt{Fujisawa_Eriguchi_2013}).
Moreover, most of the $j_\varphi /c$ (thick solid line) 
for solutions with $\kappa_0 = 4.0 $ and $\kappa_0 = 7.0$
flows oppositely against the corresponding $\mu$ current.
Since the sign of the total toroidal current determines
the sign of the magnetic flux function
(see equation \ref{Eq:GS}), this implies that
the sign of $\mu_0 \Psi$
for the whole interior region
 changes from $\mu_0 \Psi > 0$ 
to $\mu_0 \Psi < 0$ at the force-free solutions.
We calculate many solutions and confirm that the 
 the sign of $\mu_0 \Psi$ 
 for the whole interior region
changes at the force-free solution.
We call 
 the current distribution for which
$\mu_0 \Psi < 0$  oppositely flowing current. 

On the other hand, the  surface toroidal currents 
in the closed field models are always oppositely flowing  
to the total toroidal currents because of the zero-flux 
boundary condition equation (\ref{Eq:closed})
and the form of the surface current equation (\ref{Eq:j_0}).

\subsection{Deep relation between the toroidal current
and the poloidal deformations of stars}

As the many previous works pointed out,
the toroidal magnetic fields 
tend to deform stellar shapes prolate, while
the poloidal magnetic fields tend to deform 
them oblate (\citealt{Wentzel_1960,Wentzel_1961};\citealt{Ostriker_Gunn_1969};
\citealt{Mestel_Takhar_1972}).
These studies used only the magnetic fields in their formulations. 
The ideal MHD system can be described by using only magnetic fields and
one does not need to mention the electrical current density at all.
In contrast, we consider both magnetic fields and current density in our calculation.
Although these two approaches are equivalent, 
it is easier to interpret results physically in terms of the current density.
This is the reason why we consider both magnetic fields and current density in this paper.
As we have seen in the Sec.3.1, the oppositely flowing toroidal current density ($\mu_0 \Psi < 0$ )
plays a key role for appearance of the large toroidal magnetic fields. 
The direction of the toroidal current seems to relate
to the stellar deformations
because the Lorentz force is a cross product 
of current density and magnetic field.
We consider the relation between the toroidal current 
and the poloidal deformation of stars in this subsection.

In our analytic models, the Lorentz force $\Vec{L}$  
is expressed using the arbitrary function $\mu(\Psi)$ as
\begin{eqnarray}
 \Vec{L} &= 
 \left(\frac{\Vec{j}}{c} \times \Vec{B} \right)
= \rho \nabla \int \mu(\Psi)\,d \Psi = \rho \mu(\Psi) \nabla \Psi \nonumber \\
&= \rho \mu_0 \D{a}{r} \sin^2 \theta \Vec{e}_r 
 + 2\rho \mu_0 \frac{a}{r} \sin \theta \cos \theta \Vec{e}_\theta.
\label{Eq:force}
\end{eqnarray}

Following \cite{Haskell_et_al_2008}, we consider the stellar 
quadrupole deformations of $N = 1$ polytropic stars.
\cite{Haskell_et_al_2008} calculated magnetic deformations of polytorpic magnetized star with 
poloidal and toroidal magnetic fields. Although they derived the general forms of the deformations 
(equation 64, 65 \& 67 in their paper), they did not show the analytical expressions of them explicitly.
They displayed only a few numerical results in Tab. 1 in their paper. By contrast, 
we show the analytical solutions of the deformation in order 
to investigate the condition for appearance of the toroidal magnetic field dominated star. 

We assume that the influence of the magnetic fields 
to the stellar structures are small and that
their effects can be treated perturbatively.
Due to the effects of the magnetic fields,   
a certain physical quantity $X(r, \theta)$ is 
assumed to be expressed as
\begin{eqnarray}
 X(r, \theta) = X(r) + \sum_{n=0}^{\infty}  \delta X^{(n)}(r) P_n(\cos \theta),
\end{eqnarray}
where $\delta X^{(n)}$ denotes a small change 
of order $O (B^2)$ of the quantity $X$ due to 
the Lorentz force.
The angular dependencies are treated by the Legendre 
polynomial expansions and the coefficient of each Legendre polynomial
is expressed as $\delta X^{(n)}(r)$.
This expansion is also applied to the Lorentz force 
as follows:
\begin{eqnarray}
 \Vec{L}(r, \theta) = \sum_{n=0}^{\infty}  \Vec{L}^{(n)}(r) P_n(\cos \theta).
\end{eqnarray}
From the perturbed equilibrium condition equations,
the following relations can be derived:
\begin{eqnarray}
 \D{\delta p^{(n)}}{r} + \rho \D{\delta \phi_g^{(n)}}{r} 
+ \delta \rho^{(n)} \D{\phi_g}{r} = L^{(n)}_r,
\end{eqnarray}
\begin{eqnarray}
 \delta p^{(n)} + \rho \delta \phi_g^{(n)}   
= r L^{(n)}_\theta \ .
\end{eqnarray}
Since we are interested in the quadruple deformation,
we consider only $n=2$ components of Lorentz force as follows:
\begin{eqnarray}
  L_r^{(2)}
   & = &  - \frac{2 \rho \mu_0}{3} \D{a(r)}{r} \cr
  L_{\theta}^{(2)} 
   & = &  
    - \frac{2 \rho \mu_0}{3} \frac{a(r)}{r} \cr
 L^{(2)} & \equiv & 
      L_r^{(2)} -  {d (r L_{\theta}^{(2)}) \over dr} \cr
  & = &  \frac{2 \mu_0}{3}\D{\rho}{r} a(r).
\label{Eq:L_r}
\end{eqnarray}
The change of the stellar surface to the order
of the quadrupole term  can be expressed as
\begin{eqnarray}
r_d(\theta) = r_s \{1 +  \varepsilon P_2 (\cos \theta)\} 
= r_s \left\{ 1 + \frac{\varepsilon}{2}
(3\cos^2 \theta -1 )\right\},
\end{eqnarray}
where $r_d(\theta)$ denotes the deformed surface radius
and $\varepsilon$ is a small quantity which 
represents the fraction of the stellar surface
along the pole.
Following this expression, the stellar shape is prolate 
for $\varepsilon > 0$ and 
oblate for $\varepsilon < 0$.

\subsubsection{Deformation of $N \ne 0$ polytrope}

 \begin{figure*}
\begin{center}
 \includegraphics[width=8cm]{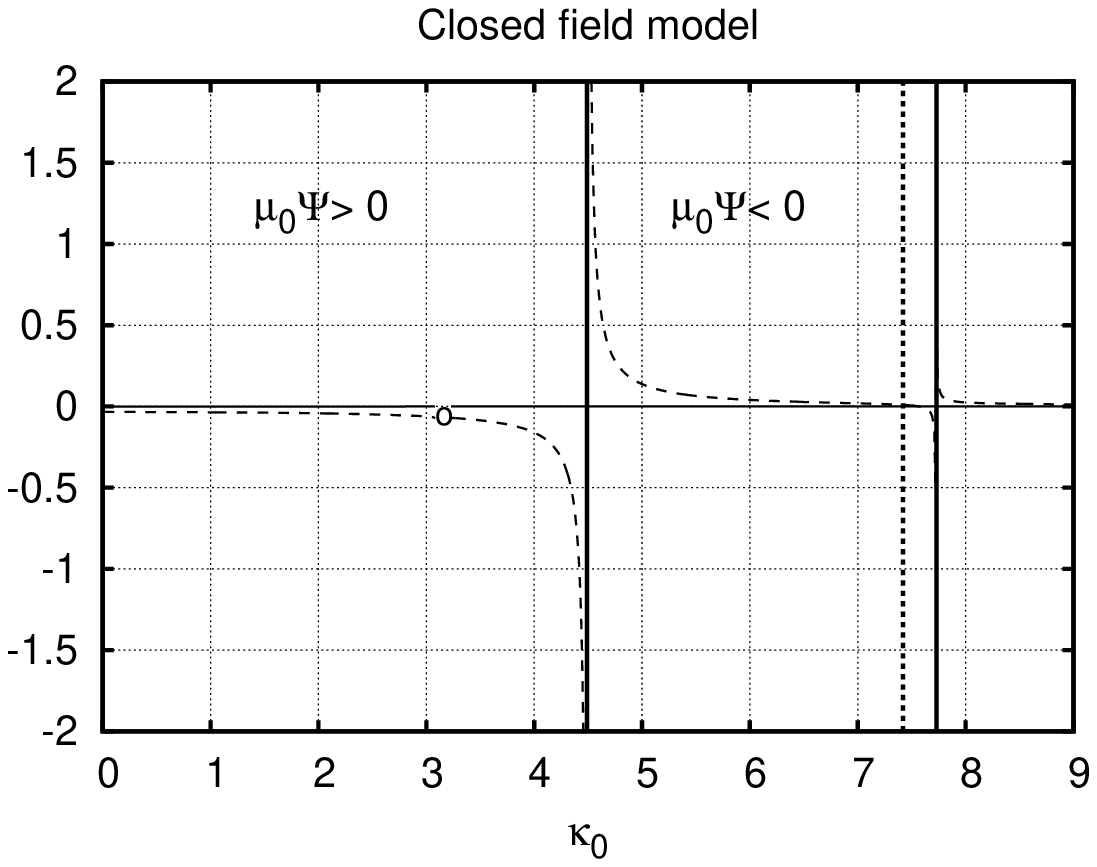}
 \includegraphics[width=8cm]{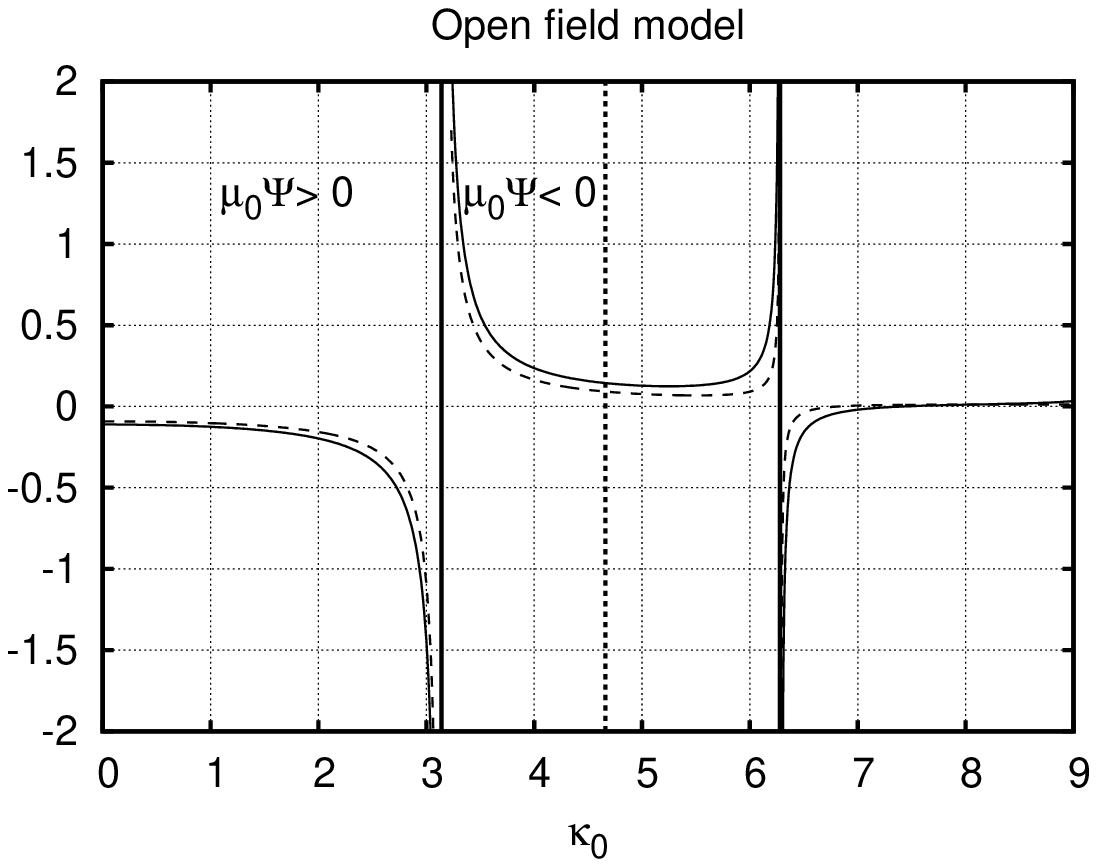}
\end{center}

\caption{The values of $-2 \mu_0 a(x=\pi)/3$ (thin solid line) and $-\delta \phi_g^{(2)}(x=\pi)$ (thin dashed line) 
in closed field model (left panel) and open field model (right panel) are plotted.
The thick vertical lines denotes force-free limit. The 
toroidal current densities consist of oppositely
flowing flows  beyond the dashed thick vertical lines.
We set $\mu_0 = -1$ and $\rho_c = 1$ in order to plot these graphs.
}
\label{Fig:eps}
 \end{figure*}

\begin{figure*}
 \begin{center}
 \includegraphics[width=4cm]{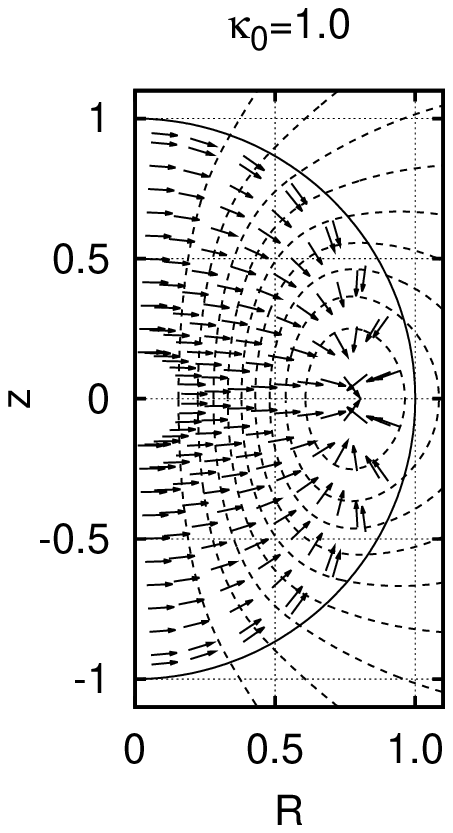}
 \includegraphics[width=4cm]{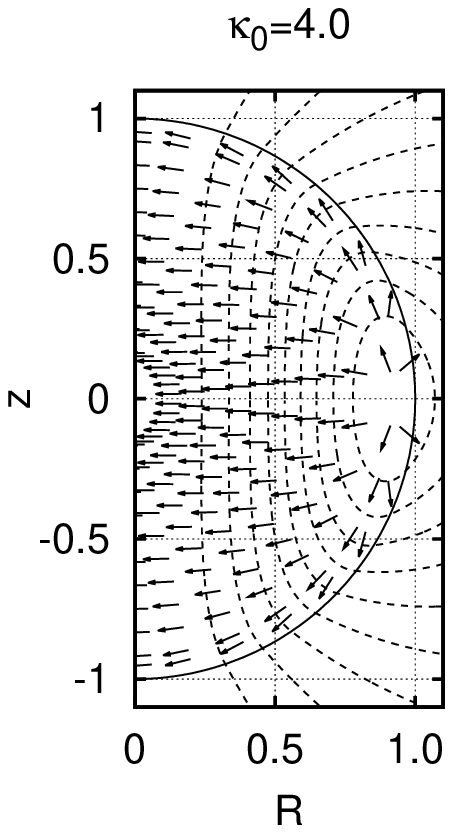}
 \includegraphics[width=4cm]{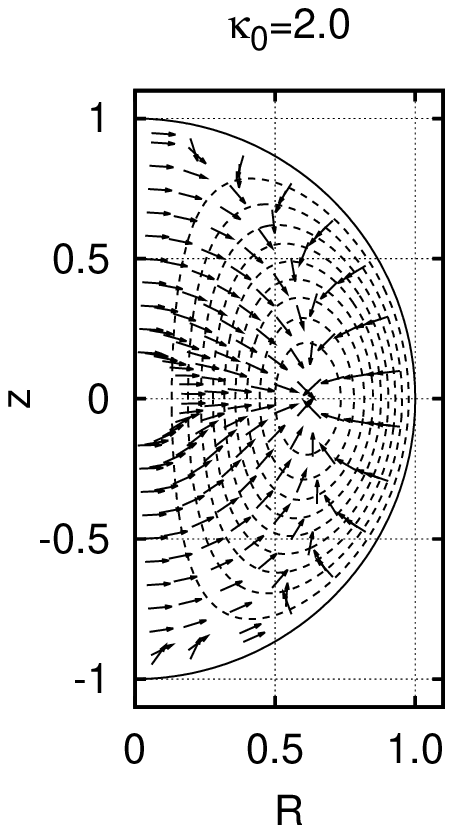}
 \includegraphics[width=4cm]{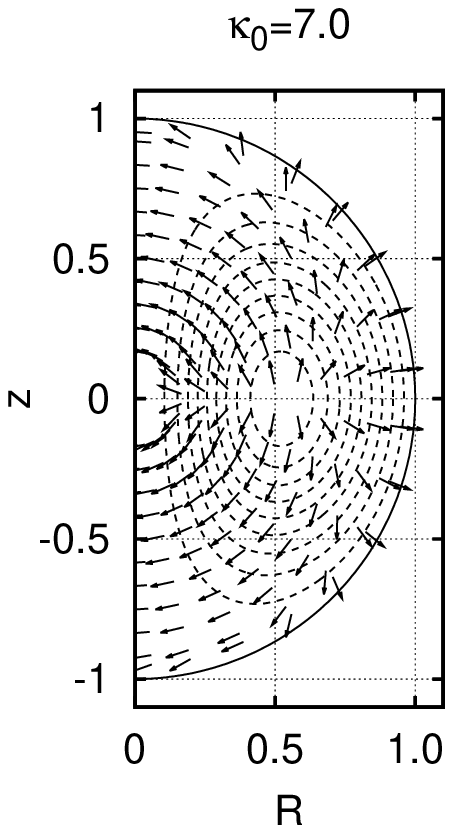}
 \end{center}

\caption{Poloidal magnetic field structures (dashed curves) and 
Lorentz force vector fields (arrows) for the open field 
models ($\kappa_0 = 1.0$, $\kappa_0 = 4.0$)
and the closed field models ($\kappa_0 = 2.0$ and $\kappa_0 = 7.0$) are 
displayed.Vectors only show their directions 
but are not scaled to their absolute values.
}
\label{Fig:force}
\end{figure*}

Using these equations, the quadrupole change of the density is 
described by:
\begin{eqnarray}
 \label{Eq:density-perturbation}
 \delta \rho^{(2)}  = \left( \D{\rho}{r} 
\delta \phi_g^{(2)} + L^{(2)}  \right) 
\left(\D{\phi_g}{r}\right)^{-1} \ .
\end{eqnarray}

Since the surface of the deformed star is defined by
a set of points where the pressure vanishes, i.e.
\begin{eqnarray}
 \label{Eq:surface-condition}
  p(r_d(\theta)) 
 =  \delta p(r_s) 
    + \varepsilon r_s P_2(\cos \theta) 
 {d p \over dr} \Big|_{r = r_s}
= 0 \ ,
\end{eqnarray}
we can derive 
\begin{eqnarray}
 r_s \varepsilon \D{\rho_0}{r} \Big|_{r=r_s} 
+ \delta \rho^{(2)} \Big|_{r=r_s} = 0 \ ,
\end{eqnarray}
for polytropes with $N \ne 0$. For $ N = 0 $
polytrope, this equation is reduced to trivial
relation $ 0 = 0$ and so we will treat $N = 0$
polytrope differently as will be shown in the 
next section.
Therefore, the quadrupole surface deformation 
$\varepsilon$  for $ N \ne 0$ is obtained by
\begin{eqnarray}
 \varepsilon = - \left(\D{\rho}{r} \right)^{-1} \frac{\delta \rho^{(2)}}{r_s} \Big|_{r=r_s}.
\end{eqnarray}
It is clearly seen that, since $\D{\rho}{r} < 0$ 
at the surface,  the stellar deformation is 
prolate for $\delta \rho^{(2)} > 0$ and  
oblate for $\delta \rho^{(2)} < 0$. In our situation,
the explicit form of $\delta \rho^{(2)}$ can be 
expressed as
\begin{eqnarray}
 \delta \rho^{(2)}
  &  = & \D{\rho}{r} \left(\frac{2 \mu_0}{3} a(r_s) 
            + \delta \phi_g^{(2)}(r_s) \right) 
\left(\D{\phi_g}{r} \right)^{-1}\Big|_{r=r_s} \ ,
\label{Eq:delta_rho}
\end{eqnarray} 
and $\varepsilon$ for $N \ne 0$ polytropes becomes as
\begin{eqnarray}
 \varepsilon 
  & = &  - \left(\frac{2 \mu_0}{3}a(r_s)
                  + \delta \phi_g^{(2)}(r)\right)
        \left(\D{\phi_g^{(2)}}{r} \right)^{-1}\Bigg|_{r=r_s} \cr
  & = & \rho \left( \D{p}{r} \right)^{-1}
     \left(\frac{2 \mu_0}{3}a(r_s)
        +\delta \phi_g^{(2)}(r)\right)
         \Bigg|_{r=r_s} \ .
\end{eqnarray}
As shown in Appendix \ref{App:N1} the gravitational change 
for $N = 1$ polytrope can be obtained as
\begin{eqnarray}
 \delta \phi_g^{(2)} (x) 
   =  \frac{F^{(p)} (x)}{x^3} 
 - \frac{1}{\pi^2}\D{F^{(p)} (\pi)}{x}\Bigg|_{x = \pi} j_2(x) \ .
\end{eqnarray}
Thus for $x = \pi$, i.e. on the surface,
\begin{eqnarray}
 \delta \phi_g^{(2)} (\pi) 
   =  \frac{F^{(p)} (\pi)}{\pi^3} - \frac{3}{\pi^4} \D{F^{(p)}}{x} \Bigg|_{x = \pi} \ ,
\end{eqnarray}
where $j_2 (\pi) = 3/\pi^2$ is used.
Here the function $F^{(p)}(x)$ is defined in Appendix \ref{App:N1}.
This is a analytic solution of the deformation.

Since the expression for the function $F^{(p)}$ is so 
complicated, it is not clearly seen the sign of the
quantity $(\delta \phi_g^{(2)}(r_s) + 2 \mu_0 a(r_s)/3)$
which determines the sign of the quantity $\varepsilon$.
In figure \ref{Fig:eps} we show the behaviors of
$-\delta \phi_g^{(2)}(r_s)$ and $-2\mu_0 a(r_s)/3$
against the value of $\kappa_0$.
As we have seen, the sign of $\mu_0 \Psi$ changes at the force-free solution 
$\kappa_0 \sim 4.49$ for closed model and $\sim \pi$ for open model.
As seen in this figure, the shape change from the
effect due to the gravitational change is the same as
that from the Lorentz term. Thus the sign of the
quantity $\varepsilon$ is essentially determined
by the sign of the Lorentz term, i.e., the sign
of the quantity $\mu_0 a(r_s)$.
Since $\rho(r)( dp /dr)^{-1} < 0$,
the stellar shape is oblate 
for 
 $\mu_0 \Psi(r,\theta) > 0$
for the whole interior region
 and prolate for 
$\mu_0 \Psi(r, \theta) < 0$
for the whole interior region
as far as the global poloidal magnetic field is 
dipole.  
Therefore, the direction of the deformation 
by Lorentz force is determined by the direction of 
the non-force free current $\mu$ current (equation \ref{Eq:j_phi}). If the 
$\mu $current flows oppositely to the 
magnetic flux ($\mu_0 \Psi < 0$), the stellar shape is 
prolate. If the $\mu$ current flows same direction ($\mu_0 \Psi > 0$),
the stellar shape becomes oblate one.
This is a deep relation between the direction of the 
toroidal current and the poloidal deformations of stars.

In figure \ref{Fig:force}, the contours of $\Psi$ 
(dashed curves) and the directions of Lorentz 
force vectors (arrows) are displayed. It should be noted that
directions of the Lorentz forces are totally
opposite between models with $\mu_0 \Psi (r_s, \theta) > 0$ ($\kappa_0 = 1.0$ and 2.0) 
and those with $\mu_0 \Psi (r_s, \theta) < 0$ ($\kappa_0 = 4.0$ and 7.0)

\subsubsection{Deformation for $N = 0$ polytrope}

For $N = 0$ polytrope, the gravitational change and
the shape change are written as follows as shown in Appendix \ref{App:N0}:
\begin{eqnarray}
 \delta \phi_g^{(2)} = -\frac{4}{5} \pi G \rho_0 
\varepsilon r^2 \ ,
\end{eqnarray}
and
\begin{eqnarray}
 \varepsilon &=&
\left\{  - \frac{4}{3}\pi G \rho_0 r_s^2  
- \left(- \frac{4}{5}\pi G \rho_0 r_s^2 
   \right)\right\}^{-1} \frac{2\mu_0}{3} a(r_s) \nonumber \\
&=& - \frac{5\mu_0}{4 \pi G \rho_0} a(r_s).
\end{eqnarray}
Here we use the stationary condition
\begin{eqnarray}
 \delta p^{(2)} + \rho_0 \delta \phi_g^{(2)} 
= r L_\theta^{(2)} \ ,
\end{eqnarray}
and the surface condition equation (\ref{Eq:surface-condition}).

Thus the sign of the
quantity $\varepsilon$ is exactly determined
by the sign of the Lorentz term, i.e., the sign
of the quantity $\mu_0 a(r_s)$.
The stellar shape is oblate 
for 
$\mu_0 \Psi(r,\theta) > 0$
 for the whole interior region
and prolate for 
$\mu_0 \Psi(r, \theta) < 0$ 
 for the whole interior region
as far as the global poloidal magnetic field is 
dipole. 
The relation between the direction of the 
$\mu$ current and the poloidal deformations of stars 
is still valid in this case.

\subsection{Deep relation between the toroidal current 
and the strong toroidal magnetic fields}

\begin{figure*}
 \begin{center}
 \includegraphics[width=8cm]{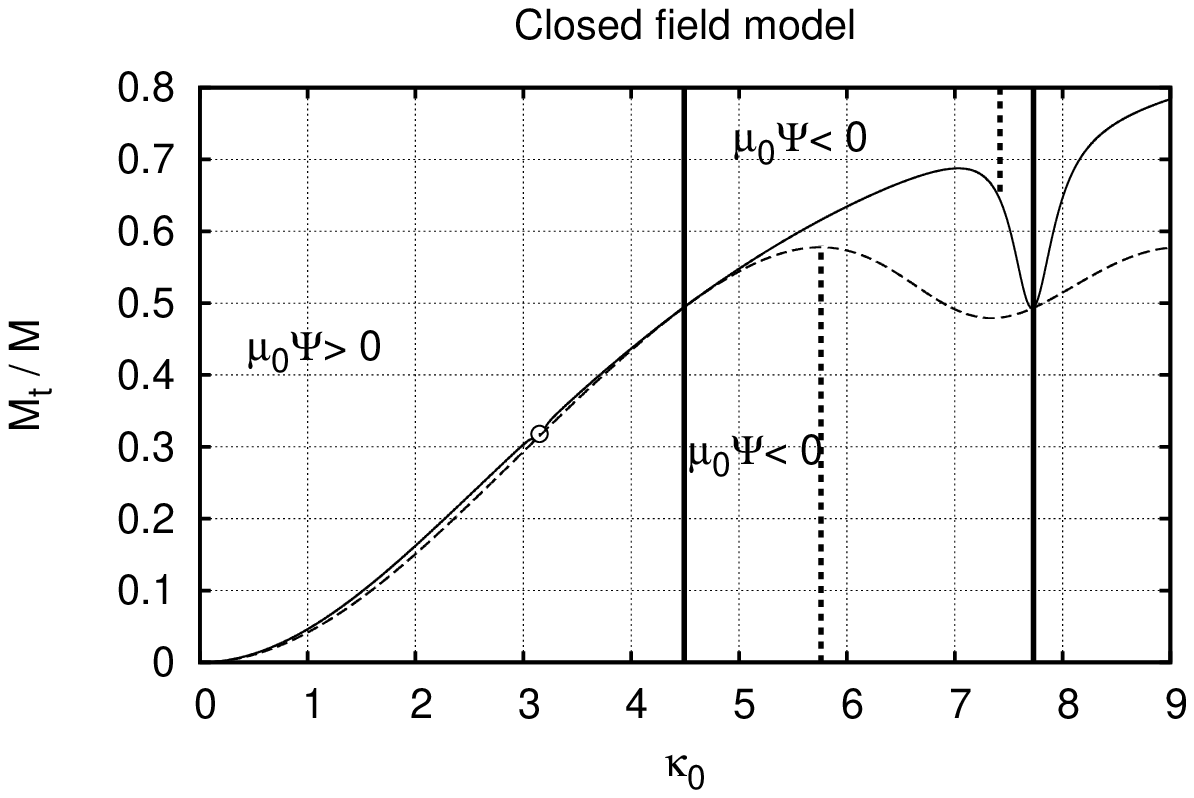}
 \includegraphics[width=8cm]{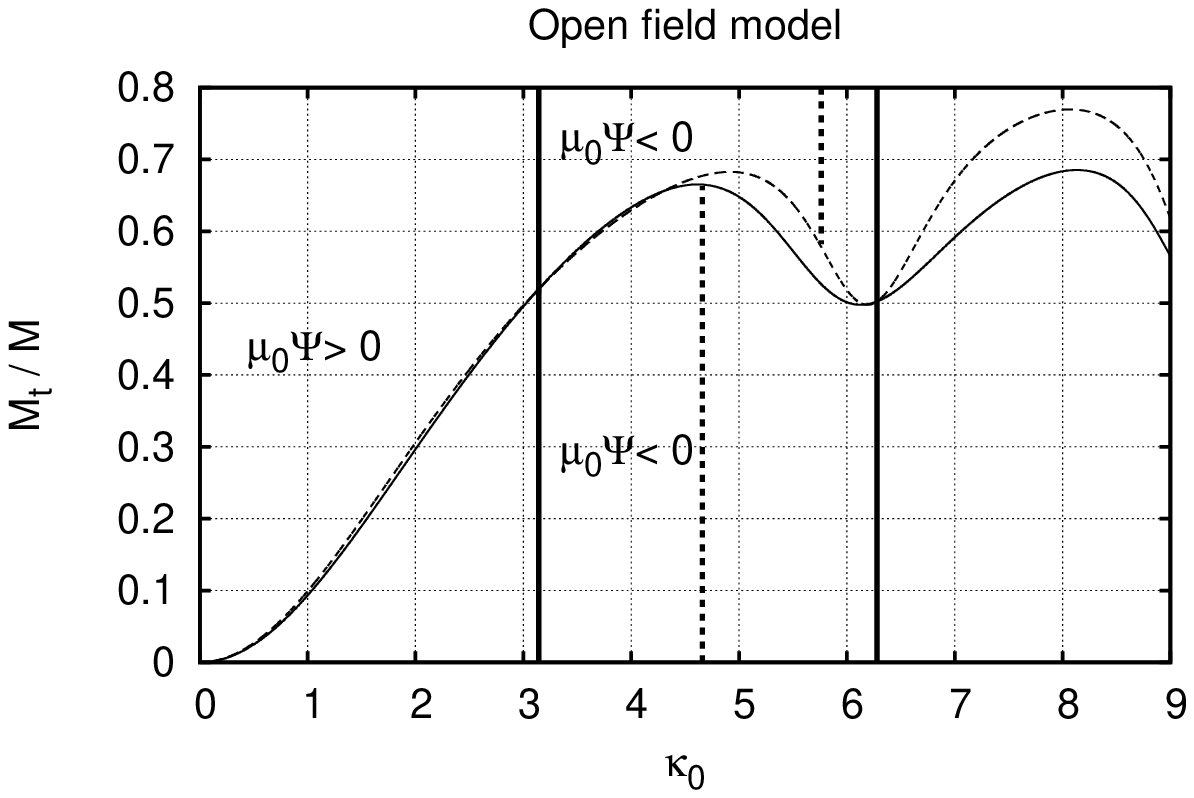}
 \end{center}

\caption{Energy ratio ${\cal M}_t / {\cal M}$ is
plotted against the value of $\kappa_0$.
Closed (left panel) and open (right panel) field 
solutions are shown. The solid and dashed curves denote 
$N=1$  and $N=0$ solutions, respectively. The vertical 
solid lines mean force-free solutions: Closed force-free
solutions appear at $\kappa_{0} \sim 4.49$ and 
$\kappa_0 \sim 7.73$ and open force-free solutions at
$\kappa_0 = \pi$ and $\kappa_0 = 2\pi$.
The toroidal current densities are composed of two
oppositely flowing components 
beyond the vertical dashed lines: $\kappa_0 \sim 5.76$ for the 
$a_{0c}$ solution 
(dashed curve in left panel), $\kappa_0 \sim 7.42$ for the 
$a_{1c}$ solution
(solid curve in left panel), $\kappa_0 \sim 5.76$ for the 
$a_{0o}$ solution 
(dashed curve in right panel) and  $\kappa_0 \sim 4.66$ 
for the $a_{1o}$ (solid curve in right panel). 
The open circle in the left panel denotes
the singular solution for $a_{1C}(r)$.
}
\label{Fig:Mt_M}
\end{figure*}

We have found a relation between the oppositely flowing 
toroidal current density and the Lorentz force in the previous subsection.
We consider a relation between the oppositely flowing 
toroidal current and the strong toroidal magnetic fields in this subsection.

In figure \ref{Fig:Mt_M},
the ratio of the toroidal magnetic field energy ${\cal M}_t$ to 
the total magnetic field energy ${\cal M} = {\cal M}_p + {\cal M}_t$  of each model
is plotted for different situations.
The solution becomes force-free at the point
denoted by the vertical solid lines.
The dashed vertical lines denote the critical values
beyond  which there arise oppositely flowing
$\kappa$ current structures dominated.

As seen in figure \ref{Fig:Mt_M}, 
$N=0$ solutions 
and $N=1$ solutions cross at $k_0 \sim 4.49$ and 
7.73 for closed field models and 
$k_0 = \pi$ and  $2\pi$ for open field models,
because the solutions at these points are force-free 
solutions as mentioned before. 
The energy ratio is ${\cal M}_t / {\cal M} \sim 0.5$
when the solutions are the first force-free
configurations. 
Therefore the solutions are divided into 
two types at the force-free solution. 
The solution whose $\kappa_0$ value is smaller than
force-free $\kappa_0$ is  poloidal dominant configuration, 
while the solution with larger $\kappa_0$ is toroidal
dominant configuration.
Since the sign of $\mu_0 \Psi$ changes at the force-free solution, 
the solution is poloidal dominant for 
$\mu_0 \Psi(r, \theta) < 0$ 
for the whole interior region
(oppositely flowing current) and the 
solution is toroidal dominant 
for $\mu_0 \Psi(r, \theta) > 0$ 
 for the whole interior region.
The oppositely flowing non-force free current 
($\mu_0 \Psi(r, \theta) < 0$ 
for the whole interior region) is required for large toroidal 
magnetic fields. 
This is a relation between the toroidal current density and 
the toroidal magnetic field.

  \subsection{A situation for appearance
of toroidal magnetic field dominated configurations}

As we have shown in previous parts in this paper, 
there are two deep relations between toroidal current, poloidal deformation and 
strong toroidal magnetic field.
One is a relation between the toroidal current 
and the poloidal deformation of stars in Sec. 3.2.
The other is a relation between the toroidal current and the strong toroidal magnetic fields. 
The important finding in this paper is 
that the appearance of oppositely flowing non force-free current 
 which fulfills the condition $(\mu_0 \Psi < 0 )$
changes the stellar shape prolate 
and  makes the toroidal magnetic fields toroidal dominant.
Therefore, a well-known relation between toroidal dominant magnetic fields 
and prolate shapes requires the oppositely 
flowing non force-free toroidal current density.
Although our result is very simple and natural, 
nobody have described explicitly that the oppositely flowing 
non force-free current density makes the stellar shape prolate.
 It might be because almost all previous studies treated only magnetic fields
and did not pay special attention to current density.

Consequently, we can conclude that
a  condition for appearance  of 
 prolate configurations and the
toroidal magnetic field dominated configurations is 
that the arbitrary function $\mu(\Psi)$ satisfies the 
following condition:
\begin{eqnarray}
 \int \mu(\Psi) d \Psi < 0 \ ,
\label{Eq:condition}
\end{eqnarray}
for the whole interior region
when the functional forms are equations (\ref{Eq:mu0}) and (\ref{Eq:I0}).
Although, exactly speaking, these analyses and conditions 
are valid within the present parameter settings, our 
results would be useful for more general situations.
This might be naively seen from the contribution of the
term $\int \mu d\Psi$ in the stationary condition 
(equation \ref{Eq:stationary-condition}).
If this term is negative, it implies that 
the action of the Lorentz term
is opposite to that of the centrifugal force which
is expressed by the term $\int \Omega(R)^2 R dR$ and 
is always positive.
In other words, the magnetic forces or Lorentz forces act
as if they are the  'anti'-centrifugal forces
and therefore shapes of stationary 
configurations become prolate
(see also calculations in \citealt{Fujisawa_Eriguchi_2014}).

Although the  condition presented 
in this paper might not be always correct, we could 
obtain the large toroidal magnetic 
fields by employing this criterion for more complicated 
calculations.

\section{Discussion and Summary }

\subsection{
Physical reason for the necessity of
appearance of $\kappa$ currents to 
realize prolate configurations}

\begin{figure*}
  \begin{center}
   \includegraphics[width=8cm]{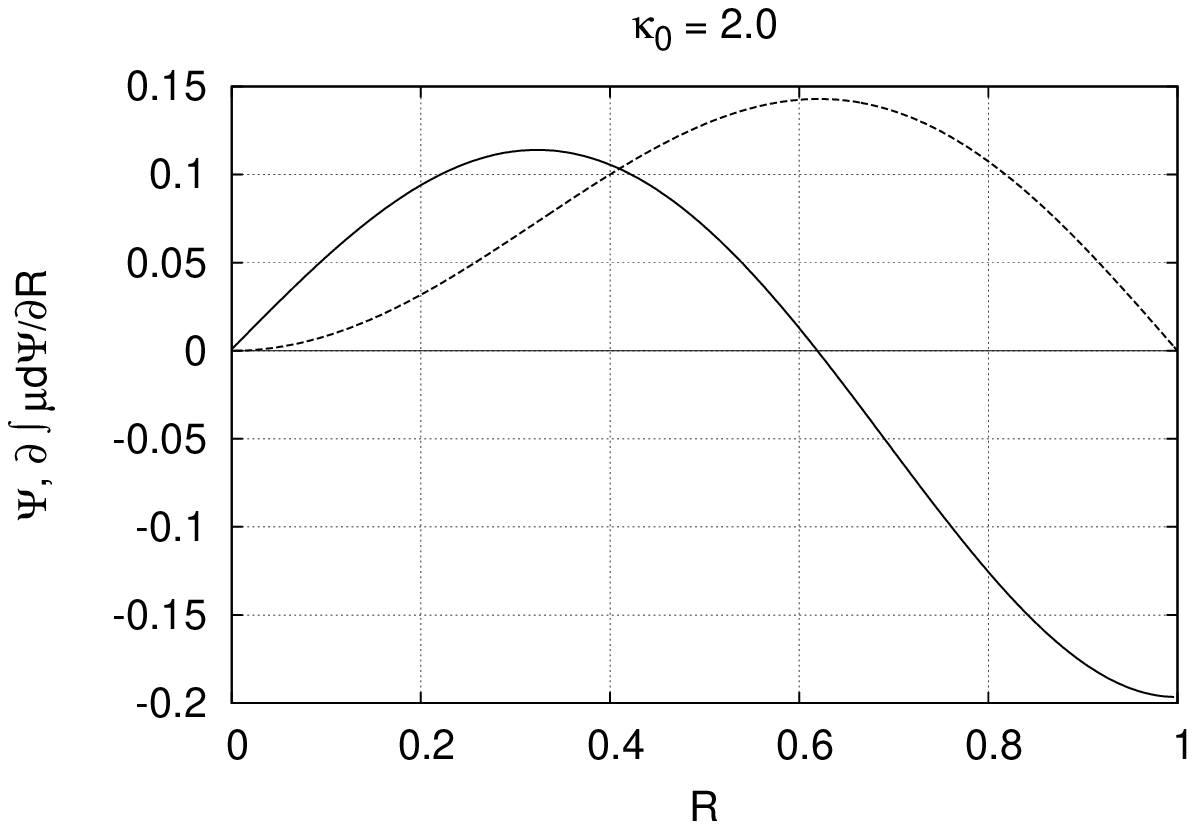}
   \includegraphics[width=8cm]{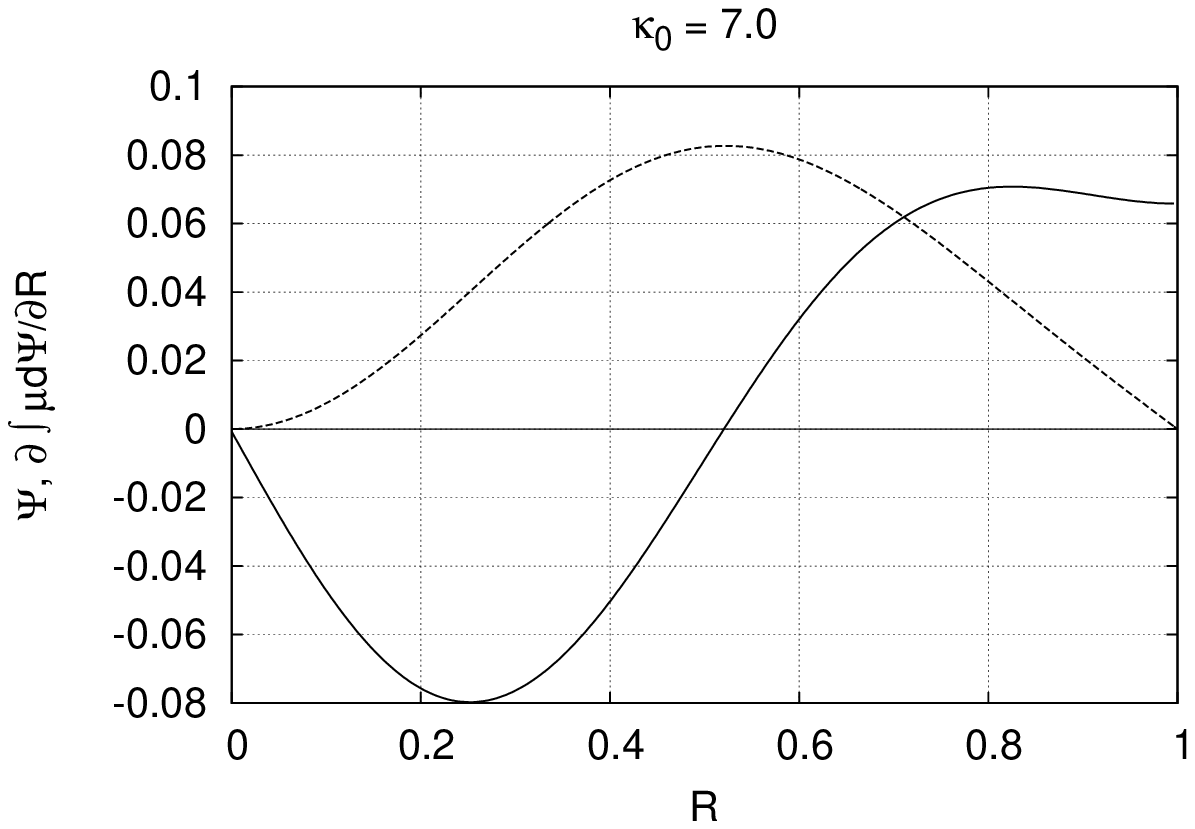}
  \end{center}
\caption{Distributions of the $\Psi$ (dashed line) and 
$\partial \int \mu d \Psi / \partial R$ (solid line) of closed field solutions 
are plotted. The left panel shows  the distributions with $\kappa_0 = 2.0$ and $\mu_0 = 1.0$
and the right panel shows those with  $\kappa_0 = 7.0$ and $\mu_0 = -1.0$.
}
\label{Fig:dPsi}
\end{figure*}

In order to get configurations with prolate shapes,
we need to include the 'anti'-centrifugal effects or
'anti'-centrifugal potentials. As is easily understood,
the anti-centrifugal potentials should behave as decreasing
functions from the symmetric axis, or, at least, they must
contain decreasing branches which cover wide enough regions
to result in effectively anti-centrifugal actions.

For our formulation, the following property is commonly 
found:
\begin{eqnarray}
  \mu > 0 \quad \rightarrow \quad 
  j^{\mu}_{\varphi} > 0 \quad \rightarrow \quad \Psi >0 
\quad \rightarrow \quad \int \mu d\Psi > 0 \ ,
\end{eqnarray}
and
\begin{eqnarray}
  \mu < 0 \quad \rightarrow \quad 
  j^{\mu}_{\varphi} < 0 \quad \rightarrow \quad \Psi < 0 
\quad \rightarrow \quad \int \mu d\Psi > 0 \ .
\end{eqnarray}
In addition to these behaviors, for $\Psi > 0$ 
configurations, 
the magnetic flux functions increase to the maximum values 
as the distance from the axis and turn to decrease beyond 
the maximum point as follows:
\begin{eqnarray}
  {\partial \int \mu d \Psi \over \partial R} & > & 0 \qquad \mbox{for} 
\quad R < R_{\max} \ , \\ 
  {\partial \int \mu d \Psi \over \partial R} & < & 0 \qquad \mbox{for} 
\quad R > R_{\max} \ ,
\end{eqnarray}
where $R_{\max}$ is the location of the maximum point
of the magnetic flux function $\Psi$ (left panel in figure \ref{Fig:dPsi}).

For $\Psi < 0$ configurations, the magnetic flux functions decrease to the minimum values as the distance
from the axis and turn to increase beyond the minimum point:
\begin{eqnarray}
  {\partial \int \mu d \Psi \over \partial R} & > & 0 \qquad \mbox{for} 
\quad R < R_{\min} \ , \\ 
  {\partial \int \mu d \Psi \over \partial R} & < & 0 \qquad \mbox{for} 
\quad R > R_{\min} \ ,
\end{eqnarray}
where $R_{\min}$ is the location of the minimum point
of the magnetic flux function $\Psi$.

Although there exist decreasing branches for both situations,
these decreasing branches cannot overcome the centrifugal 
effects due to the increasing branches. Therefore, the global
configurations with {\it purely $\varphi$-currents}
would become oblate shapes.

From this consideration, the anti-centrifugal forces could
be realized if the following (necessary) condition is
fulfilled:
\begin{eqnarray}
   \mu > 0 \quad {\footnotesize \mbox{AND}} \quad
  j_{\varphi} < 0 \quad {\footnotesize \mbox{AND}} \quad \Psi <0 
  \quad {\footnotesize \mbox{AND}} \, \int \mu d\Psi < 0 \ ,
\end{eqnarray}
or
\begin{eqnarray}
   \mu < 0  \quad {\footnotesize \mbox{AND}} \quad 
  j_{\varphi} > 0 \quad {\footnotesize \mbox{AND}} \quad \Psi >0 
  \quad {\footnotesize \mbox{AND}} \, \int \mu d\Psi < 0 \ .
\end{eqnarray}
These conditions could be realized only by 
including {\it the $\kappa$-currents}
so that the following conditions are satisfied:
\begin{eqnarray}
  \mu > 0 \ , \
  j^{\kappa}_{\varphi} < 0 \ , \ 
  j^{\mu}_{\varphi}  > 0 \ , \ 
  j_{\varphi} = j^{\kappa}_{\varphi} + 
  j^{\mu}_{\varphi}  < 0  \ 
{\footnotesize \mbox{AND}} \ \ \Psi < 0 \ ,
\end{eqnarray}
\begin{eqnarray}
  {\partial \int \mu d \Psi \over \partial R} & < & 0 \qquad \mbox{for} 
\quad R < R_{\min} \ , \\ 
  {\partial \int \mu d \Psi \over \partial R} & > & 0 \qquad \mbox{for} 
\quad R > R_{\min} \ ,
\end{eqnarray}
or
\begin{eqnarray}
   \mu < 0 \ , \
  j^{\kappa}_{\varphi} > 0 \ , \
  j^{\mu}_{\varphi}  < 0 \ , \
  j_{\varphi} = j^{\kappa}_{\varphi} + 
  j^{\mu}_{\varphi}  > 0  \
{\footnotesize \mbox{AND}} \ \ \Psi > 0 \ ,
\end{eqnarray}
\begin{eqnarray}
  {\partial \int \mu d \Psi \over \partial R} & < & 0 \qquad \mbox{for} 
\quad R < R_{\max} \ , \\ 
  {\partial \int \mu d \Psi \over \partial R} & > & 0 \qquad \mbox{for} 
\quad R > R_{\max} \ .
\end{eqnarray}
The right panel in figure \ref{Fig:dPsi} shows the distributions of $\Psi$ and 
$\frac{\partial \int \mu d \Psi}{\partial R}$ with $\kappa_0 = 7.0 $ and $\mu = - 1$. 
As seen in figures \ref{Fig:j_r} and \ref{Fig:dPsi}, the conditions of the above-mentioned  
are satisfied undoubtedly. Therefore, the appearance of $\kappa$ currents
$j^{\kappa}_{\varphi}$
which are oppositely flowing with respect to the 
$\mu$ currents $j^{\mu}_{\varphi}$  and
at the same time whose magnitudes are large enough
to overcome the $\mu$ currents are required
to realize prolate shapes.

\subsection{Twisted-torus configurations with large 
toroidal magnetic fields}

Almost all previously carried out investigations
for magnetized equilibrium states having 
twisted-torus magnetic fields had 
failed to obtain toroidal magnetic field dominated
(${\cal M}_t > {\cal M}_p$) models. We have found that 
most models of their works do not satisfy the 
condition of equation (\ref{Eq:condition}) and 
the magnetized stellar shapes are oblate
due to the $\mu$ current term. The $\kappa$
term in those works has been chosen  
as follows:
\begin{eqnarray}
 \kappa(\Psi) = \kappa_0 (\Psi - \Psi_{\max})^{k_1 + 1} \Theta(\Psi - \Psi_{\max}),
\end{eqnarray}
where, $k_1$ is a constant and  $\Theta$ is the 
Heaviside step function and $\Psi_{\max}$ is the 
maximum value of $\Psi$ on the last closed field 
line within the star.
Since the current density of this functional form
vanishes at the stellar surface, there exist no surface current 
and exterior current density.
This functional form was used by 
\cite{Tomimura_Eriguchi_2005} for the first time
and results in the twisted-torus configurations.
The same choice for the $\kappa$ has been
employed by many authors 
(e.g. \citealt{Yoshida_Eriguchi_2006}; \citealt{Yoshida_Yoshida_Eriguchi_2006};
\citealt{Kiuchi_Kotake_2008}; \citealt{Lander_Jones_2009}; \citealt{Ciolfi_et_al_2009}; \citealt{Ciolfi_et_al_2011};
\citealt{Fujisawa_Yoshida_Eriguchi_2012};\citealt{Glampedakis_Andersson_Lander_2012};
\citealt{Lander_Andersson_Glampedakis_2012}; \citealt{Fujisawa_et_al_2013};
\citealt{Fujisawa_Eriguchi_2013}; \citealt{Lander_2013a, Lander_2014}). 
While the functional form $\mu(\Psi) 
= \mu_0$ (constant) has been used in many 
investigations,
\cite{Fujisawa_Yoshida_Eriguchi_2012} and \cite{Fujisawa_et_al_2013}
used a different functional form as
\begin{eqnarray}
 \mu(\Psi) = {\mu_0}(\Psi + \epsilon)^{m},
\end{eqnarray}
where $m$ and $\epsilon$ are positive constants. 
They have obtained highly localized poloidal magnetic field 
configurations
using this type of  functional from. However, their 
works  did not satisfy the condition of equation (\ref{Eq:condition}) 
and did not 
obtain models with large toroidal magnetic fields.

Recently, \cite{Ciolfi_Rezzolla_2013} have 
adopted  a perturbative approach and
succeeded in obtaining magnetized equilibrium states 
with twisted-torus magnetic fields whose
toroidal fields are large. 
Their functional form of $\kappa$ is 
\begin{eqnarray}
 \kappa(\Psi) = \kappa_0 \Psi(|\Psi/\Psi_{\max}| - 1) 
\Theta(|\Psi/\Psi_{\max}| - 1).
\end{eqnarray}
On the other hand, the functional form of $\mu$ is 
\begin{eqnarray}
 \mu(\Psi) = c_0 [ ( 1 &-& |\Psi / \Psi_{\max}| )^2 
\Theta (1- |\Psi / \Psi_{\max}|) - \bar{k} ] \nonumber \\ 
 &+& X_0 \kappa(\Psi) \D{\kappa(\Psi)}{\Psi},
\end{eqnarray}
where $c_0$, $\bar{k} (>0) $ and $X_0$ are constants.
The toroidal magnetic field is confined within the last closed field line
in these functional forms. Outside the toroidal magnetic field region, 
the function $\kappa$ vanishes and $\mu$ becomes
\begin{eqnarray}
 \mu(\Psi) = c_0 \left[(1- |\Psi/ \Psi_{\max}|)^2 - \bar{k} \right].
\end{eqnarray}
Since the first term and the second term are positive 
and negative, respectively,
this function with larger $\bar{k}$  tends to satisfy 
the condition of equation (\ref{Eq:condition}). As they 
noted, larger values of $\bar{k}$ result in
larger energy ratios ${\cal M}_t / {\cal M}$.
As the value of $k$ increases, the energy ratio
${\cal M}_t / {\cal M}$ increases and the stellar shape becomes
more prolate in general (see Tab.1 in \citealt{Ciolfi_Rezzolla_2013}).
However, they assumed that the magnetic field configuration is purely dipole
but their functional forms and toroidal current density distribution are
far from dipole one (see bottom panels of figure 2 in \citealt{Ciolfi_Rezzolla_2013}).
Non-perturbative studies with higher order components 
were unable to reproduce their results and found contradictory results
(\citealt{Bucciantini_et_al_2015}). 
We need to calculate magnetic field configurations with higher 
order components for large toroidal models 
by using non-perturbative methods in the future. 

 The condition of equation (\ref{Eq:condition}) itself is valid when a star 
is barotropic. However, the relation between oppositely flowing 
 toroidal current density and prolate share is very simple and natural
 when a star is non-barotropic. Therefore, this condition 
 is also useful for recent perturbative non-barotropic 
 solutions (\citealt{Mastrano_et_al_2011}; \citealt{Mastrano_Melatos_2012}; 
 \citealt{Akgun_et_al_2013}; \citealt{Yoshida_2013}).
 We also need to investigate non-perturbative non-barotropic magnetized
 equilibrium states in the future. 

\subsection{Summary}
\label{Sec:Summary} 

In this paper, we have obtained four analytic solutions 
with both open and closed magnetic fields
for spherical polytropes with weak magnetic fields.

Using the obtained solutions we have discussed
the  situations for which the prolate equilibrium states and
the toroidal magnetic field dominated configurations appear. 
The main finding in this paper is that 
the appearance of the prolate shapes and the
toroidal magnetic field dominated states
are accompanied by the appearance of
oppositely flowing $\kappa$ currents
with respect to the $\mu$ current.
This situation seems to 
be related to the condition
for the non force-free toroidal current contribution,
i.e. $\int \mu(\Psi) d \Psi$, in the stationary state 
condition equation (\ref{Eq:stationary-condition}). 

Although the appearance of prolate shapes 
and the occurrence of toroidal magnetic field 
dominated states cannot be defined quantitatively,
the rough qualitative idea about them can be
determined by checking the sign of the
magnetic field potential, i.e. the quantity
$\int \mu(\Psi) d\Psi$.

Of course, the analytic solutions obtained 
in this paper have been derived under
very restricted assumptions. However,
as explained in the Discussion, the concept
of the 'anti'-centrifugal actions due to the
magnetic potentials would be applied to
more general situations for the
magnetic fields.

\begin{ack}
KF would like to thank the anonymous reviewer for useful
comments and suggestions that helped us to improve this paper.
This works was supported by Grant-in-Aid for Scientific Research on Innovative Areas, 
No.24103006.
\end{ack}

\bibliographystyle{aa}

\appendix

\section{Change of the gravitational potential for 
$N = 1$ polytrope}
\label{App:N1}

The gravitational potential perturbation for
$N = 1$ polytrope is
governed by the quadrupole component
of Poisson's equation under two boundary conditions 
($\delta \phi_g^{(2)}$ is regular at $r=0$ and 
continues the external solution smoothly at $r = r_s$):
\begin{eqnarray}
 \DD{\delta \phi_g^{(2)}}{r} + \frac{2}{r} \D{\delta \phi_g^{(2)}}{r} - \frac{6}{r^2} \delta \phi_g^{(2)} = 
4 \pi G \delta \rho^{(2)} \ .
\end{eqnarray}
Considering the density perturbation expressed by 
equation (\ref{Eq:density-perturbation}), 
this equation can be rewritten as
\begin{eqnarray}
 \DD{\delta \phi_g^{(2)}}{r} + \frac{2}{r} \D{\delta \phi_g^{(2)}}{r} &+& \left(\pi^2 - \frac{6}{r^2} \right) \delta \phi_g^{(2)}
\nonumber \\
&=& 4 \pi G \left(\D{\phi_g}{r} \right)^{-1} L^{(2)} (r).
\end{eqnarray}
By introducing the new variable $x = \pi r$,
the left-hand side of the equation is reduced to
\begin{eqnarray}
 \DD{\delta \phi_g^{(2)}}{x} + \frac{2}{x} \D{\delta \phi_g^{(2)}}{x} &+&\left(1 - \frac{6}{x^2} \right) \delta \phi_g^{(2)}
\nonumber \\
&=& \frac{4 \pi G}{\pi^2} \left(\D{\phi_g}{r} \right)^{-1} L^{(2)} (r).
\end{eqnarray}
The solution to this equation can be obtained by
taking the boundary conditions into account as follows:
\begin{eqnarray}
 \delta \phi_g^{(2)} (x) = \frac{F^{(p)} (x)}{x^3} 
- \frac{1}{\pi^2}\D{F^{(p)} (\pi)}{x}\Big|_{x = \pi} 
j_2(x),
\end{eqnarray}
where
\begin{eqnarray}
 &F&^{(p)}(x) \nonumber \\ 
 = &-& \frac{2}{3} \mu_0 A_1 \frac{\pi^2}{\kappa_0^2 
(\pi^2-\kappa_0^2)^2} 
\left[  \left\{ 6 \pi^2 \kappa_0^2 
+ \left(\pi^2 -3 \kappa_0^2 \right) \kappa_0^2 x^2 \right\} 
\sin \left(\frac{\kappa_0}{\pi}x \right) \right. \nonumber \cr
&-& \left. \frac{\kappa_0}{\pi} x\left\{6 \pi^2 \kappa_0^2 +
 \left(\pi^2 - \kappa_0^2 \right) \kappa_0^2 x^2 \right\} 
\cos \left(\frac{\kappa_0}{\pi}x\right)  \right] \nonumber \\
&-& \frac{2}{3} \mu_0 A_2 \left[ \frac{1}{2} x^4 \sin x 
+ \frac{1}{6} \left(\frac{\kappa_0^2}{\pi^2}- 1  \right) 
x^5 \cos x \right] \ .
\end{eqnarray}
Here the coefficients $A_1$ and $A_2$ are defined as 
\begin{eqnarray}
 A_1 = \frac{8\pi^2 \mu_0 \rho_c}{(\kappa_0^2 - \pi^2)^2} \frac{1}{(\sin \kappa_0 - \kappa_0 \cos \kappa_0)},
\end{eqnarray}
\begin{eqnarray}
 A_2 = \frac{4 \pi \mu_0 \rho_c}{(\kappa_0^2 - \pi^2)^2},
\end{eqnarray}
for $ N = 1$ closed configurations and
\begin{eqnarray}
 A_1 =  - \frac{4 \pi^2 \mu_0 \rho_c }{\kappa_0^2 (\kappa_0^2 - \pi^2)} \frac{1}{\sin \kappa_0},
\end{eqnarray}
\begin{eqnarray}
 A_2 = \frac{4\pi \mu_0 \rho_c}{(\kappa_0^2 - \pi^2)^2},
\end{eqnarray}
for $ N = 1$ open configurations.

\section{Surface change for $N = 0$ polytrope}
\label{App:N0}

The change of the gravitational potential
due to the change of the surface, i.e. 
$\varepsilon r_s P_2(\cos \theta)$, can be
obtained by
\begin{eqnarray}
  \delta \phi_g^{(2)}(r) 
& = & - 4 \pi G \rho_0
\int_0^{\pi} d \theta' P_2(\cos \theta') 
\int_{r_s}^{r_s(1 + \varepsilon)P_2(\cos \theta')}
dr' r'^2  {r^2 \over r'^3} \nonumber \\
& = & - {4 \pi G \rho_0 \over 5} r^2 \varepsilon \ .
\end{eqnarray}
%



\end{document}